\documentclass[useAMS,usenatbib]{mn2e}
\usepackage[totalwidth=480pt,totalheight=680pt]{geometry}
\usepackage{amsmath}
\usepackage{graphicx}
\usepackage{float}
\usepackage{siunitx}

\title[Temperatures of Kepler Eclipsing Binary Stars]{A Catalogue of Temperatures for Kepler Eclipsing Binary Stars}

\author[Armstrong, G\'{o}mez Maqueo Chew, Faedi and Pollacco]{\parbox{\textwidth}{D. J. Armstrong$^1$\thanks{d.j.armstrong@warwick.ac.uk}, Y. G\'{o}mez Maqueo Chew$^{1,2}$, F. Faedi$^1$, D. Pollacco$^1$}
\vspace{0.4cm}\\
\parbox{\textwidth}{$^{1}$University of Warwick, Department of Physics, Gibbet Hill Road, Coventry, CV4 7AL, UK\\
$^{2}$Department of Physics \& Astronomy, Vanderbilt University, Nashville, TN 37235, USA}}

\newcommand{\mytilde}{\raise.17ex\hbox{$\scriptstyle\mathtt{\sim}$}}

\begin{document}
\date{Accepted . Received}

\pagerange{\pageref{firstpage}--\pageref{lastpage}} \pubyear{2002}

\maketitle

\begin{abstract}
We have combined the Kepler Eclipsing Binary Catalogue with information from the HES, KIS and 2MASS photometric surveys to produce spectral energy distribution fits to over 2600 eclipsing binaries in the catalogue over a wavelength range of 0.36 to 2.16\AA. We present primary ($T_1$) and secondary ($T_2$) stellar temperatures, plus information on the stellar radii and system distance ratios. The derived temperatures are on average accurate to 370K in $T_1$ and 620K in $T_2$. Our results improve on the similarly derived physical parameters of the Kepler Input Catalogue through consideration of both stars of the binary system rather than a single star model, and inclusion of additional U band photometry. We expect these results to aid future uses of the Kepler Eclipsing Binary data, both in target selection and to inform users of the extremely high precision light curves available. We do not include surface gravities or system metallicities, as these were found to have an insignificant effect on the observed photometric bands.
\end{abstract}

\begin{keywords}
binaries:general, binaries:close, binaries:eclipsing, Astronomical Databases: catalogues
\end{keywords}

\section{Introduction}
In preparation for the Kepler mission \citet{Brown:2011dr} produced the Kepler Input Catalog (KIC), providing spectral energy distribution fit parameters for stars in the Kepler field of view. Since then the Kepler satellite has produced high-precision light curves for some 150000 stars in this field, many of which have proven to be eclipsing binaries. These are catalogued in the Kepler Eclipsing Binary Catalogue (KEBC) \citep{Prsa:2011dx,Slawson:2011fg,Matijevic:2012di}, and number well over 2000. With this catalogue as a guide, many interesting results have been found \citep[e.g.][]{Carter:2011kx,Rappaport:2012ic,Bloemen:2012je,Armstrong:2012ie,Lee:2013ee} not least those of the circumbinary planets \citep[e.g.][]{Doyle:2011ev,Welsh:2012kl}. The parameters presented by the KIC are used in target selection for both the primary Kepler purpose of planet hunting and other guest observer programs, as well as to provide estimated radii for candidate planets \citep{Batalha:2013fg}. They have been subject to latter testing, through for example population synthesis \citep{Farmer:2013wj}. 

Here we aim to produce a catalogue similar to the KIC for the eclipsing binary systems of the KEBC, taking into account our new knowledge of their binary nature with the information presented by the various photometric surveys of the Kepler field. In this way we can improve upon the KIC for these binary systems, through extended wavelength coverage (particularly inclusion of the U band), and consideration of both stars. Although the primary star often dominates the observed flux, not including the secondary (as in the KIC) can lead to biases.

The structure of the paper is as follows. We lay out our input data in Section \ref{sectdata}, and explain the model used to fit the observed colour bands in Section \ref{sectmodel}. We test the model against simulated stellar parameters in Section \ref{secttesting}, and present our results catalogue and parameter distributions in Section \ref{sectresults}, while Section \ref{sectdiscuss} discusses the results and observed distributions.

\section{Data}
\label{sectdata}
\subsection{KEBC}
We make use of data from the Kepler Eclipsing Binary Catalogue (KEBC). Our targeted objects' KIC Identification Numbers are taken from the Catalogue, along with the primary and secondary eclipse depth ratio, which were used to produce an estimate of the temperature ratio $T_2/T_1$ of each binary as described in Sect. \ref{t2t1gen}. We took a version of the KEBC as presented online on 18-09-2013 to give KIC IDs and eclipse depth ratios, yielding 2610 systems. Thirteen of these systems contained multiple entries in the catalogue, we took the first entry only in each of these cases. Specific KIC IDs used are presented with our results in Section \ref{sectresults}.
 
\subsection{HES}
\label{sectHES}
We use photometry from the Howell-Everett Survey (HES)\citep{Everett:2012kq}. The survey consists of the three optical filters Johnson U B and V, in the Vega system \citep{Morgan:1953hy}, and contains data on 2424 objects of the 2610 from the KEBC. Errors were taken as presented in the HES catalogue.

\subsection{KIS}
\label{sectKIS}
In parallel to the Howell-Everett Survey, we use data from the Kepler INT Survey (KIS) \citep{Greiss:2012ii,Greiss:2012uq}, Data Release 2. This allows models to be fit to two independent sets of photometry separately, increasing reliability and allowing bad data to be more easily flagged. The KIS provides data in the RGO U, Sloan g, r and i bands, in the Vega system. Errors provided in the catalogue are photometric only, we add systematic errors to the photometric errors in quadrature. We use Table 3 of \citet{Greiss:2012ii}, which gives the systematic offset used in calibrating each band to the KIC, as an estimation of the systematic error for each band (a 0.05 mag systematic error was used for the U band).

Some objects in the KIS are observed more than once, these are available as separate sets of data, duplicates or triplicates (no object had more than 3 sets of data), for the same object. There are 2439 of the KEBC systems present in the KIS, of which 764 are duplicates and 111 triplicates. These multiple dataset systems were treated as independent objects during the subsequent analysis. Data for individual bands were filtered using the KIS class flag. Only bands of data with class -1 (stellar) or -2 (probably stellar) were used.

\subsection{2MASS}
\label{sect2MASS}
To each of the HES and KIS surveys we added data from 2MASS \citep{Skrutskie:2006hl}. This consisted of the Johnson J H and 2MASS specific Ks bands, in the Vega system. The combined total photometric uncertainties were used as presented in the 2MASS catalogue. Data were accepted if the SNR in that band was $\geq 5$ and the object was not flagged as blended or contaminated. 2590 objects were found in the 2MASS catalogue.

\subsection{Combination}
For each object the above survey data was combined to produce two partially independent datasets, HES + 2MASS, hereafter UBVJHK, and KIS + 2MASS, hereafter UgriJHK. Each dataset is used separately in what follows, allowing comparison between results derived from the HES and KIS surveys and hence increasing reliability.

\section{Model}
\label{sectmodel}
\subsection{Setup}
We use a Markov Chain Monte Carlo code utilising the Metropolis-Hastings algorithm, enacted using the python module PyMC \citep{Patil:Huard:Fonnesbeck:2010:JSSOBK:v35i04}. We assumed each system to be composed of two stars, and fit the combined contribution from these two stars to the observed colour data. This intrinsically assumes that the data were taken when the stars were out of eclipse. This is reasonable for many eclipsing binaries, for those in overcontact systems (where the two stars are permanently in contact) it is less so but hard to avoid. In these cases it is only the apparent radii of the eclipsed star which will change; this is in general fit poorly anyway (see Section \ref{secttesting}), and the fit temperatures should be unaffected.

\subsubsection{Model Atmospheres}
We use the Castelli-Kurucz 04 Model Atmospheres \citep{Castelli:2004ti}. These cover a grid of [$3500K<\textrm{T}<50000K$], [$-2.5<[M/H]<0.5$], and [$0.0<\textrm{Log g}<5.0$], of which for computing efficiency purposes we used temperatures up to 13000K (some systems were run with higher temperature limits, see Sect \ref{sectresfitting}). The grid spacing was 250K in Temperature (1000K for atmospheres above 13000K), 0.5 in $[M/H]$ (with an additional point at $[M/H]=0.2$) and 0.5 in Log g. Model values were interpolated linearly between the two closest grid points of each parameter. Although the CK atmospheres depend on surface gravity and metallicity as well as temperature, we found that they were retrieved extremely poorly (See Sect \ref{secttesting}). As such we did not include them in our model, using CK atmospheres with the KIC surface gravity and zero metallicity for each system.

To make use of the CK atmospheres, they must be integrated over a response function for the relevant filter to produce band-integrated flux densities. We used filter transmission curves as detailed in the respective papers of the HES and KIS. For the 2MASS data, relative spectral response functions from \citet{Cohen:2003gg} were used. These provide an absolute flux calibration using the calibrated spectrum of Vega, matching with the Vega system magnitudes of the HES and KIS.

\subsubsection{Interstellar Extinction}
We use the extinction relations of \citet{Cardelli:1989dp} with a constant $R_V$ of 3.1, resulting in two analytical relations relevant for optical wavelengths (U to i) and IR wavelengths (JHK). This allowed extinction in each band to be calculated as a function of that of the V-band. The specific conversion factor for each photometric band depends on the spectrum of the star under question (due to the distribution of stellar flux within the band), but we found that making the simplifying assumption of an extinction factor for each band calculated at the central wavelength of the relevant band had negligible effect for the stars considered here. We took V band extinctions from the $A_V$ values of the KIC, and used the mean value of the KIC EBs of 0.4 mag for systems where no KIC values were available. This applied to 244 systems of the 2610; these systems are flagged in the presented catalogue. While the KIC values for $A_V$ are by no means perfect (see \citet{Brown:2011dr} for a full discussion) we found that fitting them ourselves did not constrain them, and results in an additional free parameter in what is already a large parameter space. As such we use the KIC values both to constrict the parameter space and to allow easier comparison between our results and the $T_{\textrm{eff}}$ of the KIC.

\subsubsection{Generation of $T_2/T_1$}
\label{t2t1gen}
As direct values for $T_2/T_1$ were not available at the time of submission, we estimate it from the ratio of secondary to primary eclipse depth (as $T_2/T_1 \simeq (\textrm{depth}_\textrm{sec}/\textrm{depth}_\textrm{pri})^{0.25}$). For circular binaries, this represents a good proxy for the temperature ratio. For increasingly eccentric orbits, due to the possibility of different surface areas being occulted in primary and secondary eclipse, the eclipse depth ratio becomes an increasingly less accurate estimator of $T_2/T_1$. We formed a distribution for $T_2/T_1$ by including 1) The known parameters (period, eccentricity, argument of periapse) of each binary, 2) measurement scatter in recovering the eclipse depths (gaussian errors of 0.025 and 0.05 for over contact and non-overcontact binaries respectively, from the test Figures 8 and 10 of \citet{Prsa:2011dx}) and 3) a correction for the effect of eccentricity, derived for each binary individually. For full details see Appendix \ref{appt2t1gen}.

\subsubsection{Fit Parameters}
Four fit parameters were used. These comprised the primary star temperature $T_1$, secondary star temperature $T_2$ (constrained through the temperature ratio as measured from the lightcurves), radius ratio of the stars $R_2/R_1$, and the primary radius to system distance ratio $R_1/D$. Note that stellar radii as used provide no allowance for non-sphericity of stars, and as such represent an `effective radius', particularly in the case of overcontact eclipsing binary systems. No constraining relations were used, each parameter was allowed to vary according to its prior (see Section \ref{sectpriors}). Observables were treated as having normally distributed errors, and comprised each available colour band.

\subsubsection{Input Data and Priors}
\label{sectpriors}
The fits were performed to the combined UBVJHK dataset (6 colour bands), and separately to the UgriJHK dataset (7 colour bands). Each covered the wavelength range 0.36 to 2.16 \AA. Missing (not available from the relevant photometric survey catalogue) or bad as defined in Sections \ref{sectHES}, \ref{sectKIS}, and \ref{sect2MASS}) data was given an error of $10^5$ magnitudes to ensure it did not affect the fit. We treated the KEBC temperature ratio as an observable with distribution as described in Section \ref{t2t1gen}, and used this to constrain the temperature of the secondary star from that of the primary. We assumed priors on the model as detailed in Table \ref{Tablepriors}. Where no KIC $T_{\textrm{eff}}$ was available for an object, 5000K was used as the prior mean for $T_1$. We tested the effect of the prior on T1 by running 1000 of the binary systems with firstly a prior of 5000K with standard deviation 2000K, and secondly a prior of the KIC $T_{eff}$ with the same standard deviation, in each case with no extinction. The offset in the means of the $T_1$ distributions was 14K, so no significant systematic effect is caused by the prior. The standard deviation of the difference between the two cases, excluding non-converged systems, is \mytilde200K, well within the errors we quote in Section \ref{secterrors}. As such we conclude that the choice of this prior has no significant effect on the retrieved values. The `primary' star of each system was chosen using the KEBC temperature ratios - these values were taken for the purposes of fitting, even when they were greater than unity. In the final catalogue the primary star values have been set as the star with the dominant flux contribution, as calculated from the temperature and radius ratios of the model output.

\begin{table*}
\caption{Model Priors}
\label{Tablepriors}
\begin{tabular}{@{}lllllr@{}}
\hline
Parameter  & Distribution & Parameters & & &\\
  &  &  Mean& Standard Deviation & Lower Limit & Upper Limit\\
   \hline
  $T_1$ & Normal  &  KIC $T_{\textrm{eff}}$ &2000K & 3500K & 13000K (see Sect \ref{sectresfitting}) \\
  $T_2/T_1$ &See Sect \ref{t2t1gen}  & & & & \\
      $R_1/D$  &   Normal & 0.003$R_\odot\textrm{pc}^{-1}$ &0.05$R_\odot\textrm{pc}^{-1}$ & $10^{-5}R_\odot\textrm{pc}^{-1}$ & 0.2$R_\odot\textrm{pc}^{-1}$\\
   $R_2/R_1$  &   Normal & 0.8 &0.3 & 0.01 & 3.0\\   
\hline
\end{tabular}
\end{table*}

\subsection{Testing}
\label{secttesting}
The model was tested on a simulated distribution of 1000 binary systems, for both the UBVJHK and UgriJHK datasets. These systems were generated with separate distributions for each physical stellar parameter, as no complete unbiased distribution could be found for these parameters in binary stars. We used the distributions as laid out in Table \ref{Testdistributions} (with all values constrained to be above zero), which were designed to cover the expected parameter space for the Kepler mission EBs. The distribution of Temperature Ratio $T_2/T_1$ was taken as a gaussian approximation to the distribution of the KEBC. Note that this form of test involves generating fake colours using the very model atmospheres and filter transmissions used to fit them. Also while it involves realistic parameter values, these do not combine to represent `real' stars. Hence this is purely a test of information content in the used colour bands. No significant difference was seen between each dataset, as expected from their similar colour bands and wavelength ranges.

\begin{figure}
\resizebox{\hsize}{!}{\includegraphics{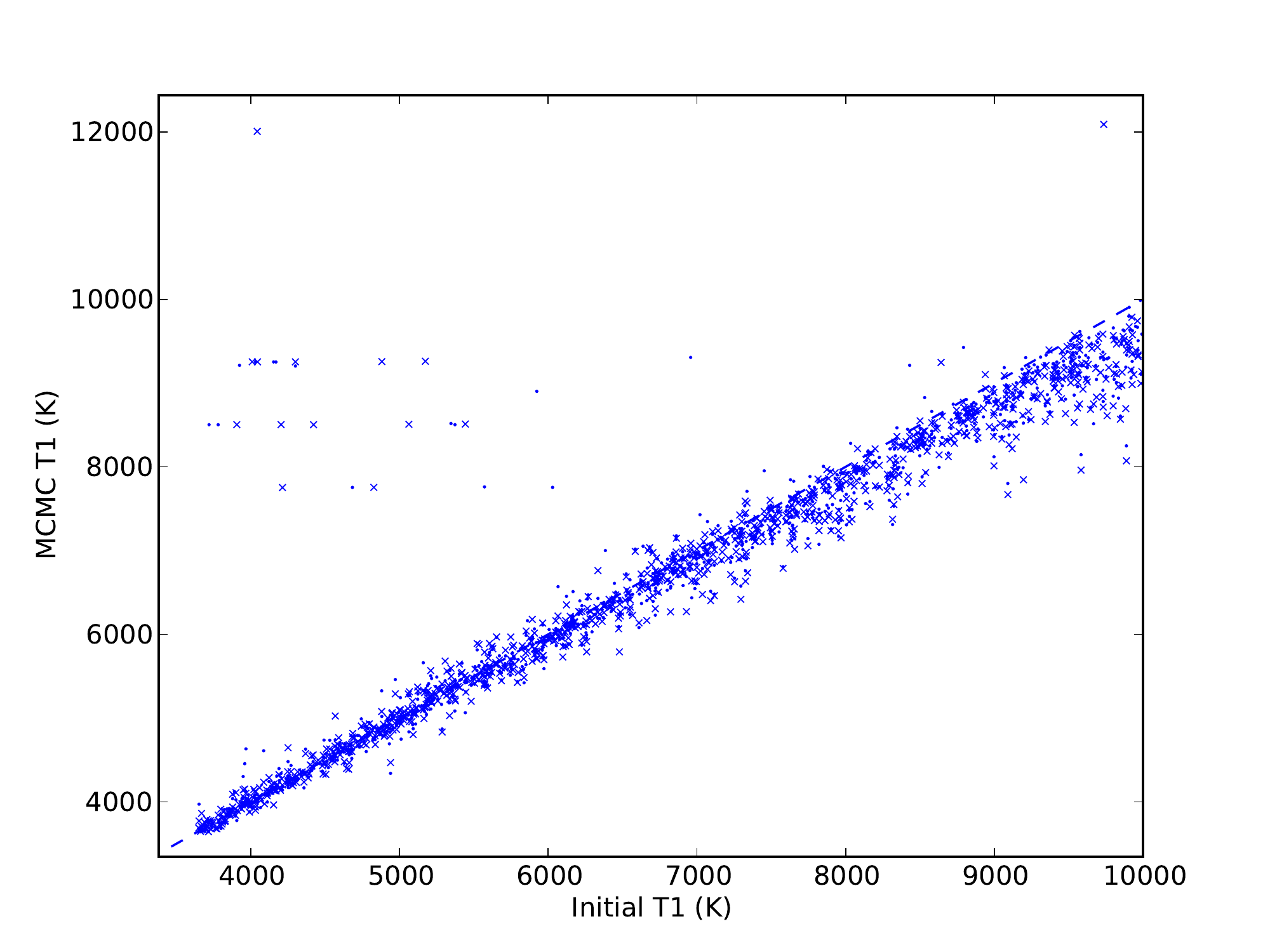}}
\caption{Input and MCMC fit primary star temperatures T1 for 1000 simulated sets of stelar parameters. The dashed line shows a perfect match. Dots represent the UBVJHK dataset and crosses the UgriJHK. The small number of systems highly deviant from the dashed line have not converged.}
\label{t1test}
\end{figure}

\begin{figure}
\resizebox{\hsize}{!}{\includegraphics{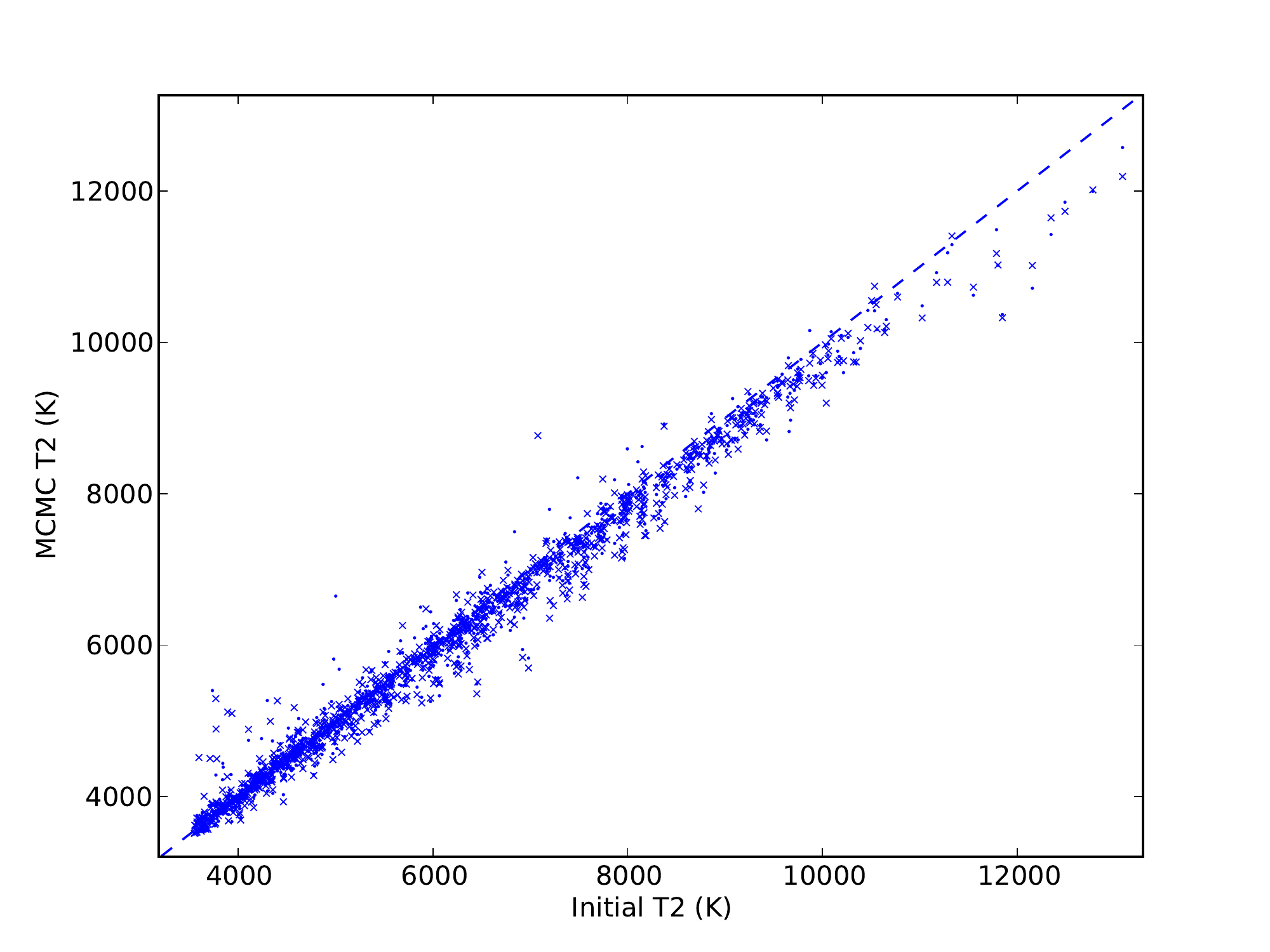}}
\caption{As Figure \ref{t1test} for T2}
\label{t2test}
\end{figure}

\begin{figure}
\resizebox{\hsize}{!}{\includegraphics{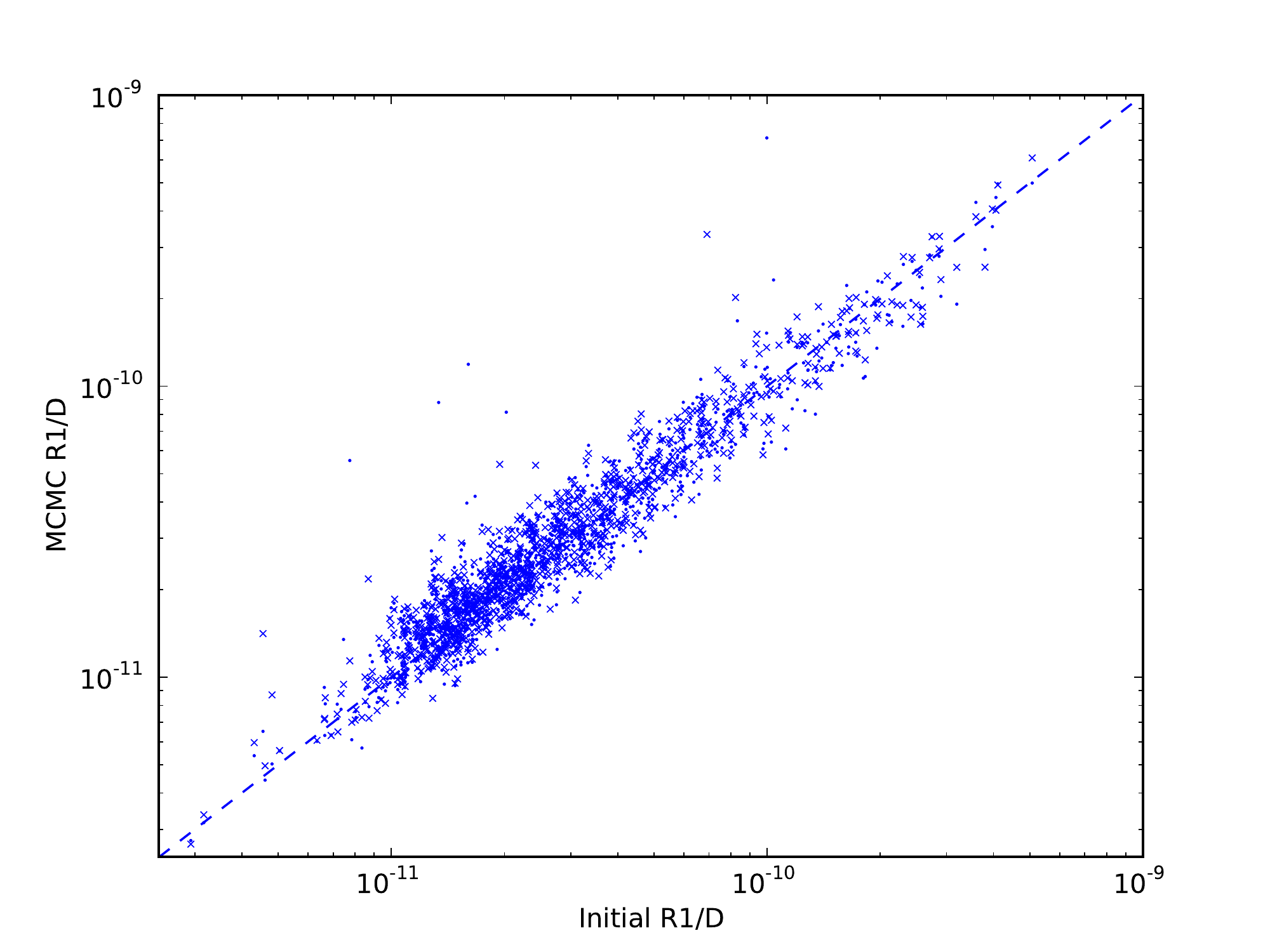}}
\caption{As Figure \ref{t1test} for $R_1/D$}
\label{r1overdtest}
\end{figure}

\begin{figure}
\resizebox{\hsize}{!}{\includegraphics{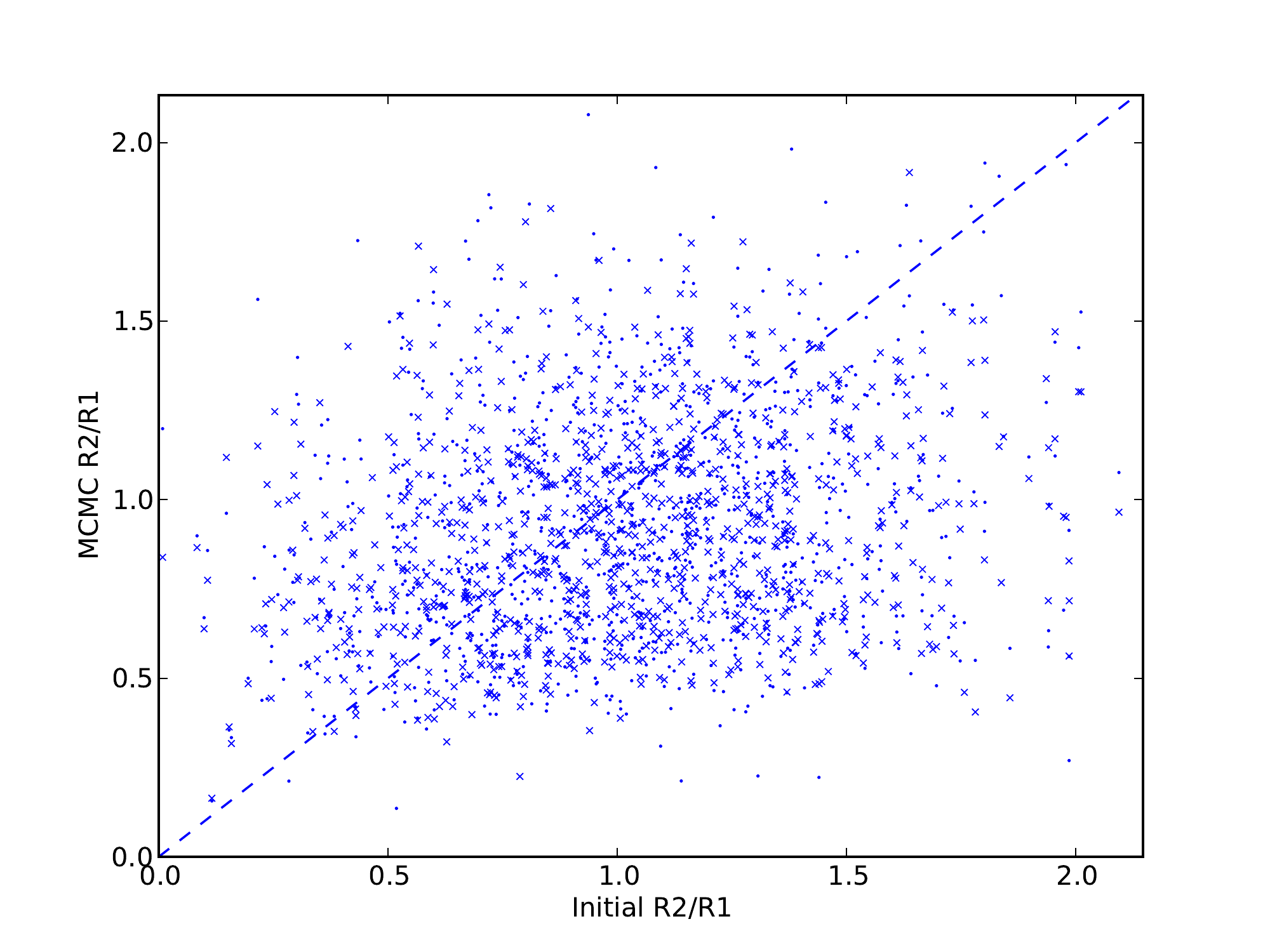}}
\caption{As Figure \ref{t1test} for $R_2/R_1$}
\label{rrattest}
\end{figure}

Simulated colour bands were generated via integrating over the CK04 model atmospheres as detailed above. The MCMC was run for 50000 iterations with a burn in period of 20000 iterations. No significant extinction was included in this test. The retrieved values of $T_1$, $T_2$, $R_1/D$ and $R_2/R_1$ as compared to their input values for each dataset are shown in Figures \ref{t1test}, \ref{t2test}, \ref{r1overdtest} and \ref{rrattest}. The agreement in all parameters except $R_2/R_1$ shows that 20000 is a sufficient number of iterations to allow convergence for the majority of systems. To remove as many as possible of those few remaining unconverged, a higher number of iterations is specified for the real data. The retrieval of surface gravity and metallicity was extremely poor. These parameters have very little impact on the observed colours within their error. As such these parameters were not included in the model. As shown above, the combination $R_1/D$ replaces $R_1$ and $D$ in the actual model run, as the latter two were not individually constrained.

\begin{table}
\caption{Distribution of Test Parameters}
\label{Testdistributions}
\begin{tabular}{@{}lllr@{}}
\hline
Parameter  & Distribution & Parameters &\\
  &  &  Mean,& Standard Deviation\\
    &   &Lower Limit, &Upper Limit\\
\hline
  $T_1$ & Uniform  &  3500K &10000K \\
 $T_2/T_1$    & Normal  &  0.9123& 0.1668 \\
     $R_1$  &   Normal & 0.8$R_\odot$ &0.2$R_\odot$\\
    $D$   &   Uniform  & 50 pc &1500 pc \\
   $R_2/R_1$  &   Normal & 1.0 &0.4\\   
   Log $g_1$  &   Normal & 4.5 cgs&0.2 cgs\\   
   Log $g_2$  &   Normal & 4.5 cgs& 0.2 cgs\\   
   $[M/H]$  & Uniform  & -2.5& 0.5 \\
   $A_V$  & Normal & 0.05 mag &0.02 mag \\
\hline
\end{tabular}
\end{table}

\begin{figure}
\resizebox{\hsize}{!}{\includegraphics{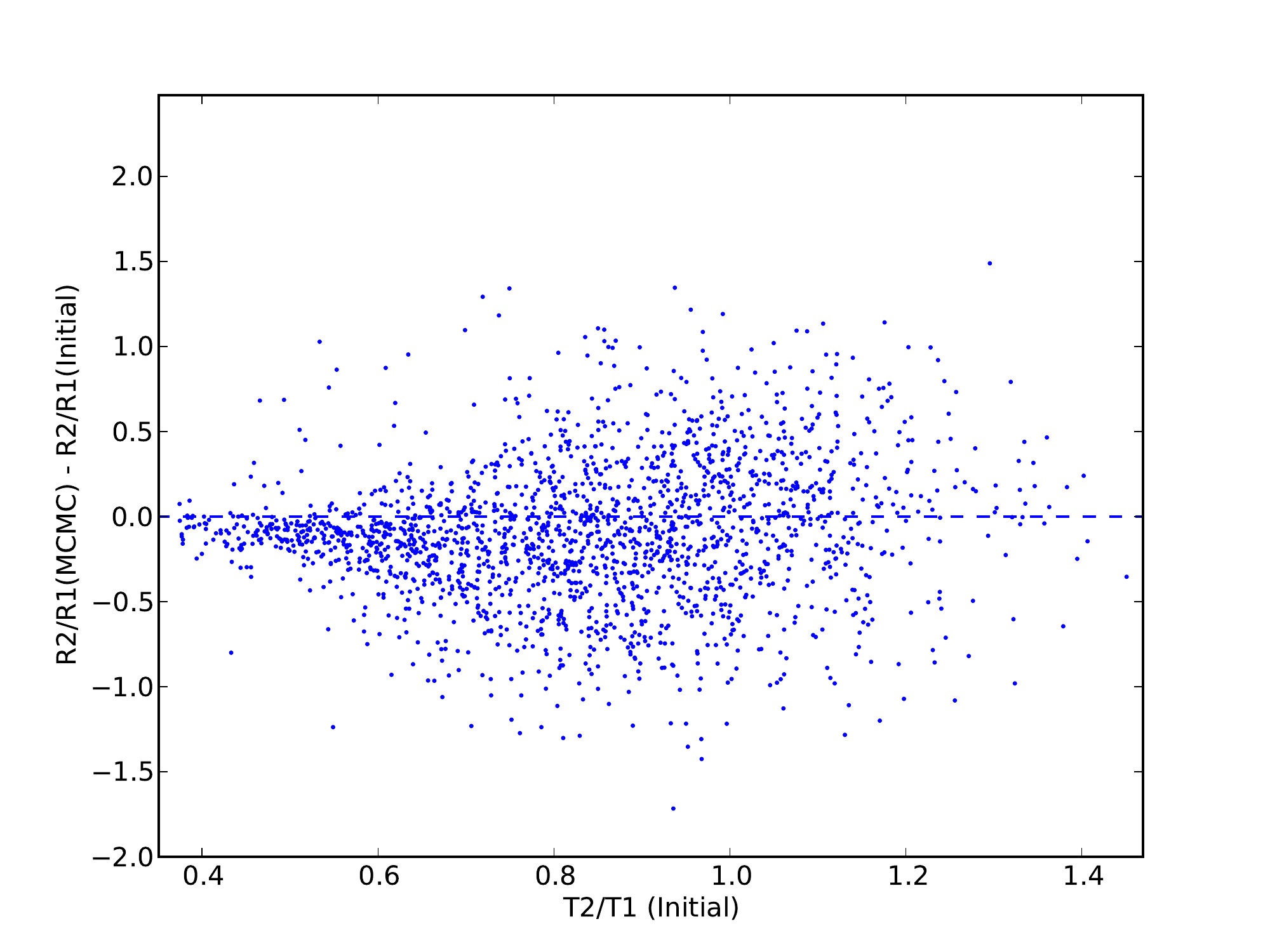}}
\caption{The dependence of $R_2/R_1$ fit quality on $T_2/T_1$. The 1000 simulated sets of parameters used in Figure \ref{t1test} plus 1000 additional sets with lower temperature ratios are shown.}
\label{rrattrattest}
\end{figure}

For two stars with well separated temperatures it should be possible to fit each stellar atmosphere and hence obtain information on both stellar temperatures and also the ratio of the radii of the stars (to each other and to the system distance). For the systems in the KEBC however, the two stellar temperatures in general proved to be too close to allow this, as the peak emission of each star is often located too close in wavelength to that of the other star. This led to the well-retrieved information being in general the primary star temperature $T_1$ (and through the temperature ratio the secondary temperature $T_2$) along with a combination of parameters which we term the binary solid angle, equal to $(R_1^2 + R_2^2)/D^2$. The relevance of the binary solid angle as opposed to $R_1/D$ and $R_2/R_1$ depends on the temperature ratio. For ratios significantly different from unity the individual components $R_1/D$ and $R_2/R_1$ become well retrieved. The difference between input and MCMC fit values for $R_2/R_1$ using the UBVJHK colours is shown as a function of $T_2/T_1$ in Figure \ref{rrattrattest} (with an additional 1000 systems with lower temperature ratios added to illustrate the correlation). Note the systematic offset of about -0.2 even at lower values of $T_2/T_1$. In what follows we publish both $R_1/D$ and $R_2/R_1$, as each are in some cases accurate, but users should note the above in choosing whether to use these values individually or combined into the binary solid angle mentioned. The relevance of $R2/R1$ should be determined from Figure \ref{rrattrattest} in line with the needs of the user. We note that when $T_2/T_1$ approaches unity that for main sequence stars $R_2/R_1$ should also be close to unity.

\section{Results}
\label{sectresults}
\subsection{Fitting Parameters}
\label{sectresfitting}
Each object was run through the MCMC for 100000 iterations, including a burn in period of 30000 iterations (significantly more than was used in the test of Section \ref{secttesting}). At 250 iteration intervals through the burn in phase, the model was tuned. Objects with KIC $T_{eff} <= 9000K$ were run using a high temperature limit of 13000K for efficiency. Other objects (including those with no KIC value for $T_{eff}$) were run with a limit of 50000K, and are flagged in the catalogue.

In forming final values for each object, output parameters were checked for consistency between both datasets, and also between duplicate or triplicate UgriJHK datasets where available. We used $T_1$ as the parameter for checks (with $T_1$ defined as the temperature of the star dominating the system flux) , as this was the strongest recovered parameter while testing. First, all fits with 3 or less bands were excluded. Then, in the cases where more than one fit was available, each fit was checked against each of the others to see if it was within $3\sigma$ (using the maximum $\sigma$ of the two fits being checked). If the fits were thus consistent, they were both included in the weighted average (using their MCMC derived errors) to calculate the final set of values. This process was repeated for all possible fit combinations (the maximum number of fits for a systems is 4, one UBVJHK and 3 UgriJHK). In this way highly deviant fits (for example due to bad photometric data) can be excluded from the final values where possible. For systems where no good fits were present, final values were formed using the weighted average of all available fits, and errors by the standard deviation of those fit values. These systems are flagged in the catalogue to highlight the systematic difference between their fits and/or the lack of photometry available.

\subsection{Errors}
\label{secterrors}
We use the MCMC derived errors to represent the gaussian noise associated with our model fitting. These have median values of 120K for $T_1$, 310K for $T_2$, \num{8.2e-5}$R_\odot/pc$ for $R_1/D$ and 0.17 for $R_2/R_1$. We also estimate the effect of extinction on the presented values, as the KIC extinctions are known to have particularly high error. We compare a run of the model on the UBVJHK dataset with the KIC extinction values (as in the final run) and with no extinction. The error on the KIC extinction values (\mytilde0.3) is of the order of the values themselves (mean \mytilde0.4), so the difference in model fits generated by removing extinction represents a reasonable estimate of the error they could cause. We found that there was a median difference between fits with and without extinction of 350K and 540K for $T_1$ and $T_2$, \num{7.3e-5}$R_\odot/pc$ for $R_1/D$, and for $R_2/R_1$ 0.15. As such, extinction has a significant effect. We treat these median extinction effects as a $1\sigma$ additional gaussian error on the presented values, and give the combined error in the catalogue. While this is an estimate, neither the KIC extinction values nor their errors are characterised enough for a more detailed approach to be meaningful.

We note that this error estimation does not take account of systematic offsets caused by contamination (`third light'), bad data, or bad extinction values, and objects affected will have larger errors. While we removed bad or contaminated photometric data where it was labelled as such, and applied errors to the input temperature ratios to reduce the impact of bad values, these issues will remain for some objects.

We then adopt $1\sigma$ errors formed from a combination of the effect of the extinction systematic and fitting noise values. Errors are presented individually for each system, but to give a guideline the median catalogue error is 370K in $T_1$, 620K in $T_2$, \num{2.5e-12} in $R_1/D$, and 0.23 in $R_2/R_1$. For high temperatures ($>9000K$) the errors on the temperature are larger, as seen in Figure \ref{t1test}) and discussed in Section \ref{shortcomings}. The errors on the radius parameters will vary with $T_2/T_1$, as described in Sections \ref{secttesting} and \ref{shortcomings}. The derived temperature errors are consistent with the temperatures of various systems in the KEBC which have been studied in detail, see Section \ref{sectdiscuss} for detail. 

\subsection{Catalogue}
The format of our results is presented in Table \ref{catformat} (the full catalogue available online). For each object the results are given for each fitted dataset, along with `final' values. These are formed from the average of the available `good' fits - fits can be excluded as explained in Section \ref{sectresfitting}. Entries are flagged for various reasons; a summary of the used flags, in order in which they appear in the catalogue, is given in Table \ref{tabflags}.

\begin{table}
\caption{Catalogue Flags}
\label{tabflags}
\begin{tabular}{@{}ll@{}}
\hline
Column & Flag Description \\
\hline
1 & 1 if object run with 50000K temperature limit\\
2 & 1 if no KIC $T_{eff}$, $A_V$ or Log g values available\\
3 & 1 if candidate or proven three-body system\\
4 & 1 if no KEBC eclipse information available\\
5 & 1 if no matching fits available with 4+ colour bands\\
\hline
\end{tabular}
\end{table}

\begin{table*}
 \centering
  \caption{Catalogue Format. UBV = UBVJHK, KIS=UgriJHK}
    \label{catformat}
  \begin{tabular}{@{}llll@{}}
  \hline
   Column Header     &  Description & Column Header  &  Description  \\
 \hline
 KIC ID & Kepler Input Catalogue Identifier & $T_2$ KIS$_1$, Error &  -1 if None\\
$T_1$ Final, Error& Final Primary Temperature and error (K)& $R_1/D$ KIS$_1$, Error  &-1 if None \\
$T_2$ Final, Error& Final Secondary Temperature and error(K)& $R_2/R_1$ KIS$_1$, Error& -1 if None \\
$R_1/D$ Final, Error & Primary Radius / System Distance and error& KIS$_1$ Bands Used  & 1 if used, order -- -- -- UgriJHK  \\
$R_2/R_1$ Final, Error & Component Radii Ratio and error& $T_1$ KIS$_2$, Error &  -1 if None \\
$T_2/T_1$ Input  &     Input prior temperature ratio   &$T_2$ KIS$_2$, Error  &  -1 if None \\
   Flags & 4 digit flag of fit quality, see Table \ref{tabflags}  & $R_1/D$ KIS$_2$, Error& -1 if None\\
Output Source  &   Final values source, UBV, KIS$_1$,KIS$_2$,KIS$_3$, 1 if used  & $R_2/R_1$ KIS$_2$, Error&  -1 if None \\
  $T_1$ UBV, Error  &   -1 if None & KIS$_2$ Bands Used&1 if used, order -- -- -- UgriJHK \\
 $T_2$ UBV, Error & -1 if None  &$T_1$ KIS$_3$, Error&  -1 if None \\
  $R_1/D$ UBV, Error &  -1 if None&$T_2$ KIS$_3$, Error &   -1 if None  \\
$R_2/R_1$ UBV, Error &    -1 if None   &$R_1/D$ KIS$_3$, Error& -1 if None\\
UBV Bands Used&  1 if used, order UBV-- -- -- -- JHK   &$R_2/R_1$ KIS$_3$, Error& -1 if None \\
$T_1$ KIS$_1$, Error & -1 if None &KIS$_3$ Bands Used&  1 if used, order -- -- -- UgriJHK \\
\hline
\end{tabular}

\end{table*}

\subsection{Distributions}
\begin{figure}
\resizebox{\hsize}{!}{\includegraphics{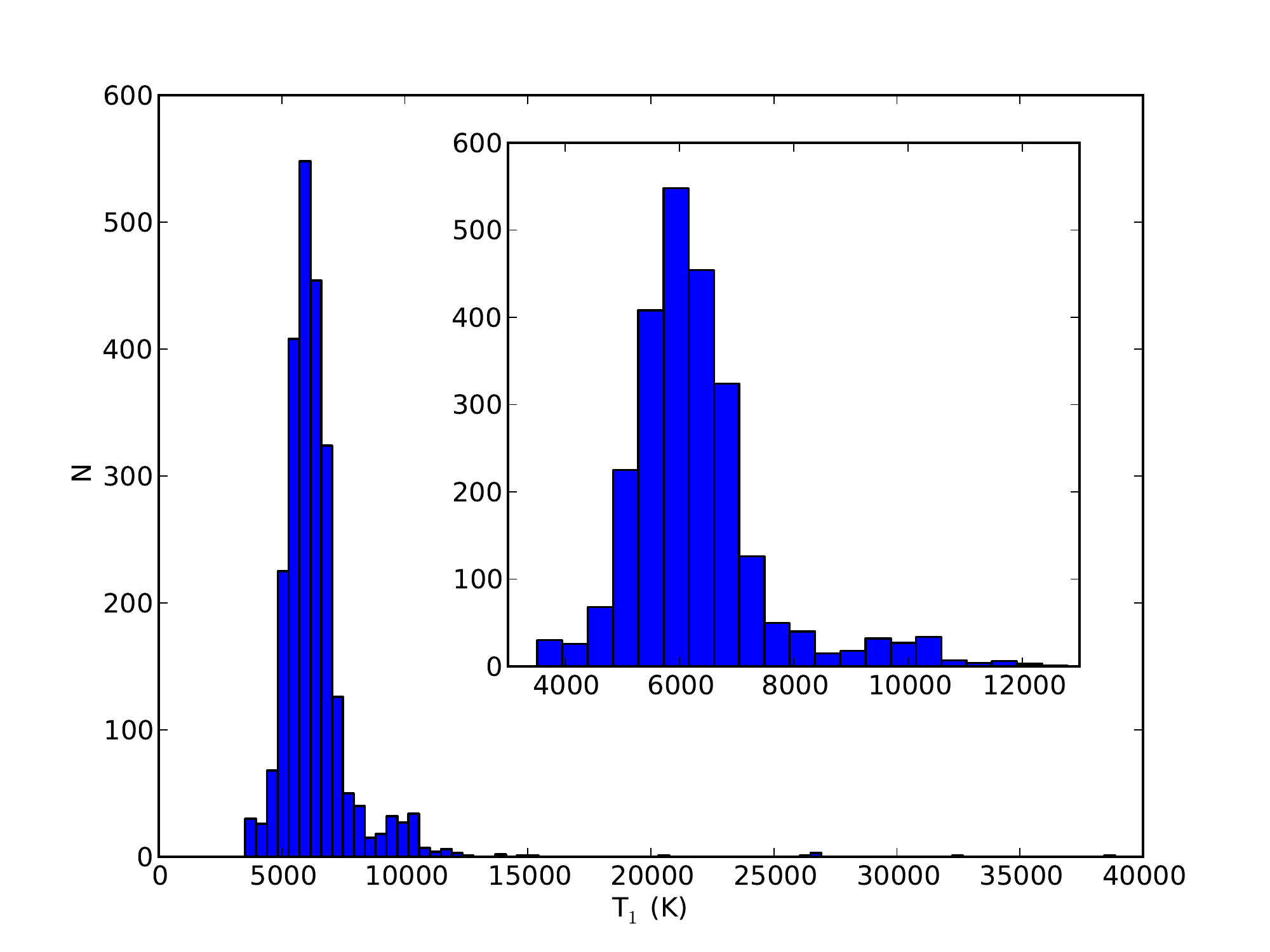}}
\caption{The distribution of primary star temperature $T_1$, drawn from the `final' catalogue values. Systems with no consistent fits or lacking 4+ photometric bands are excluded. The cooler temperatures where the majority of our sample lies are shown in the inset.}
\label{t1dist}
\end{figure}

\begin{figure}
\resizebox{\hsize}{!}{\includegraphics{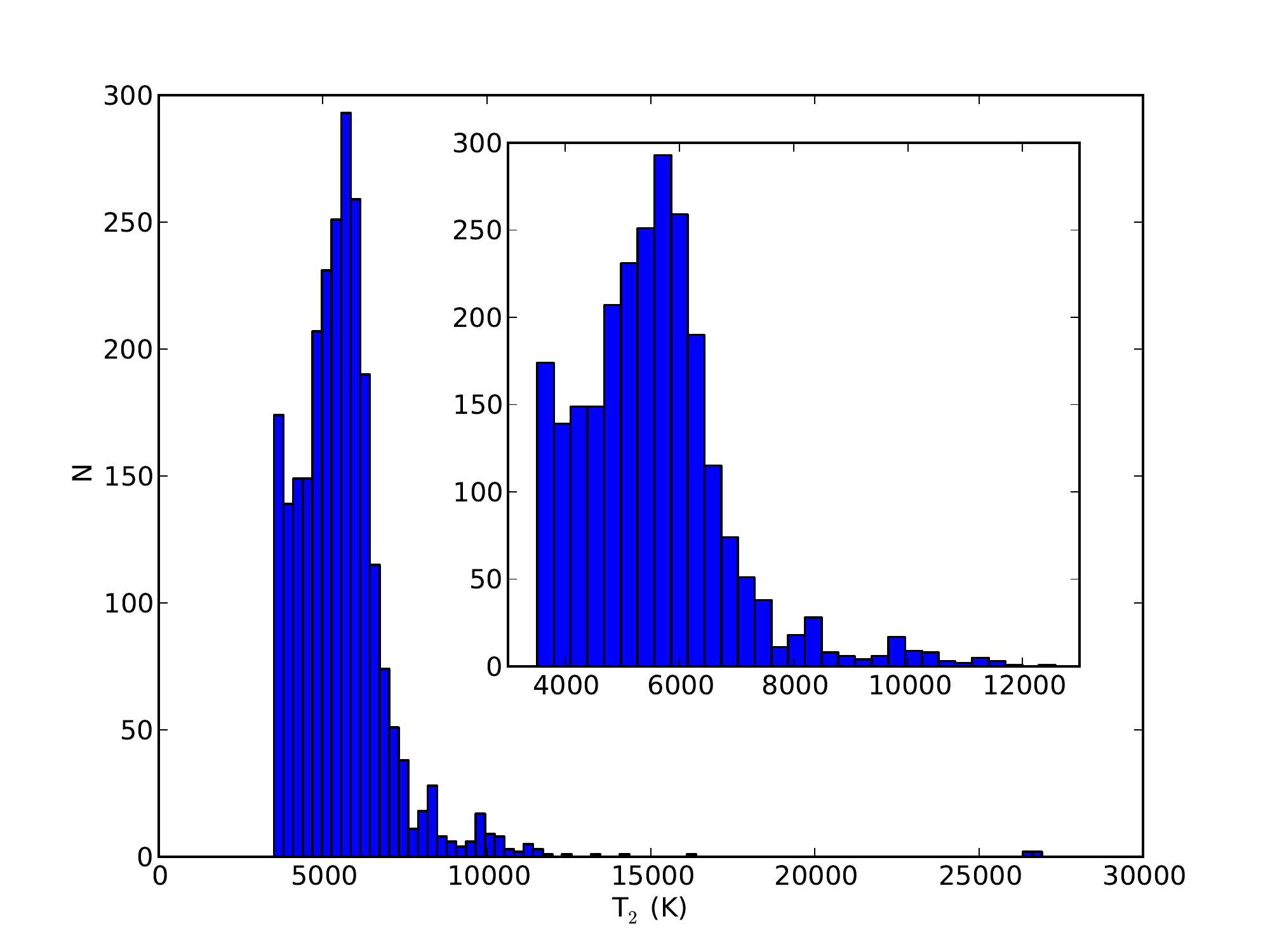}}
\caption{As Figure \ref{t1dist} for secondary star temperature $T_2$.}
\label{t2dist}
\end{figure}

\begin{figure}
\resizebox{\hsize}{!}{\includegraphics{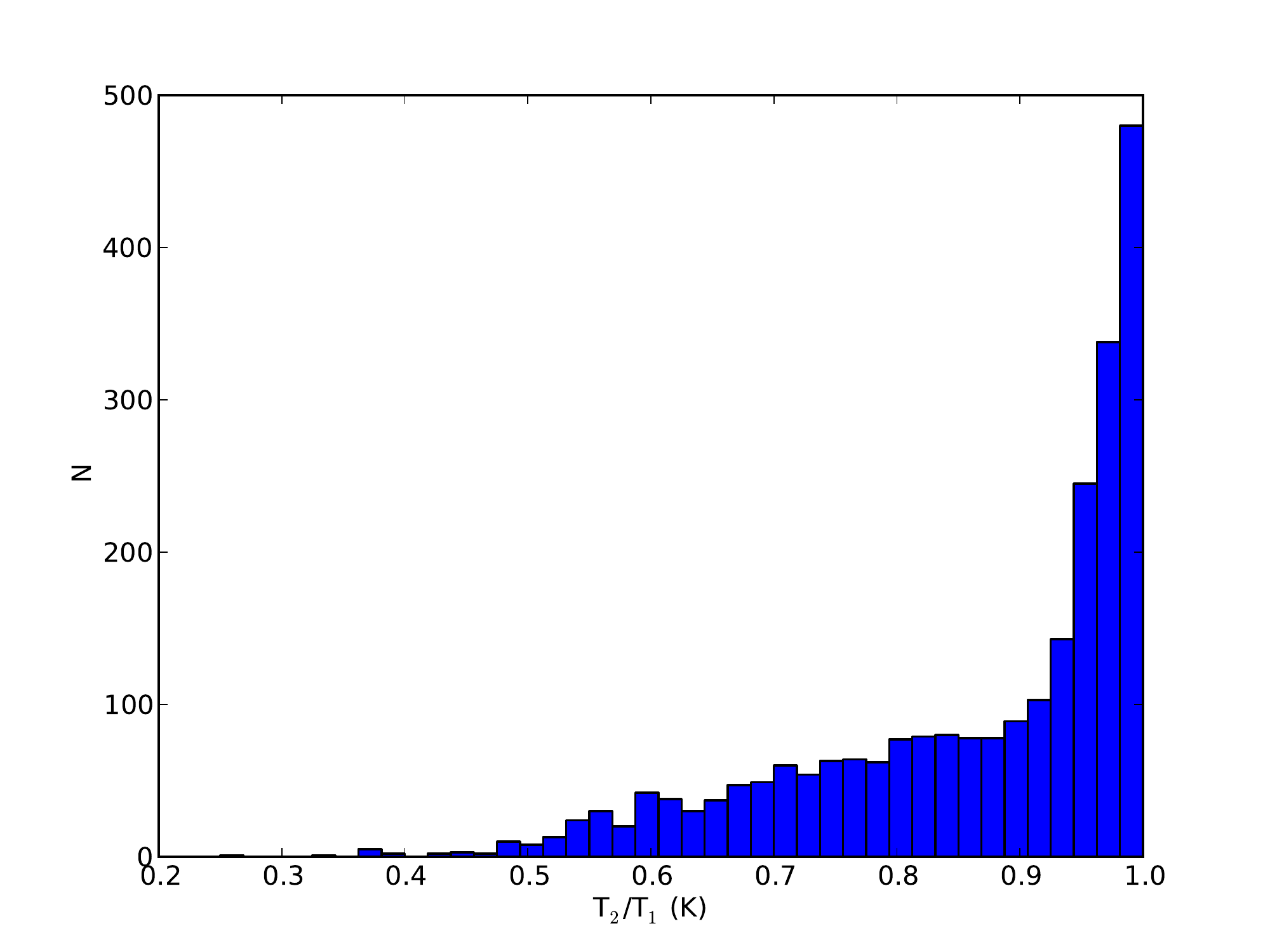}}
\caption{The total distribution of temperature ratio $T_2/T_1$, drawn from the `final' catalogue values. Systems with $T_2/T_1$ are included as their inverse.}
\label{t1t2dist}
\end{figure}

\begin{figure}
\resizebox{\hsize}{!}{\includegraphics{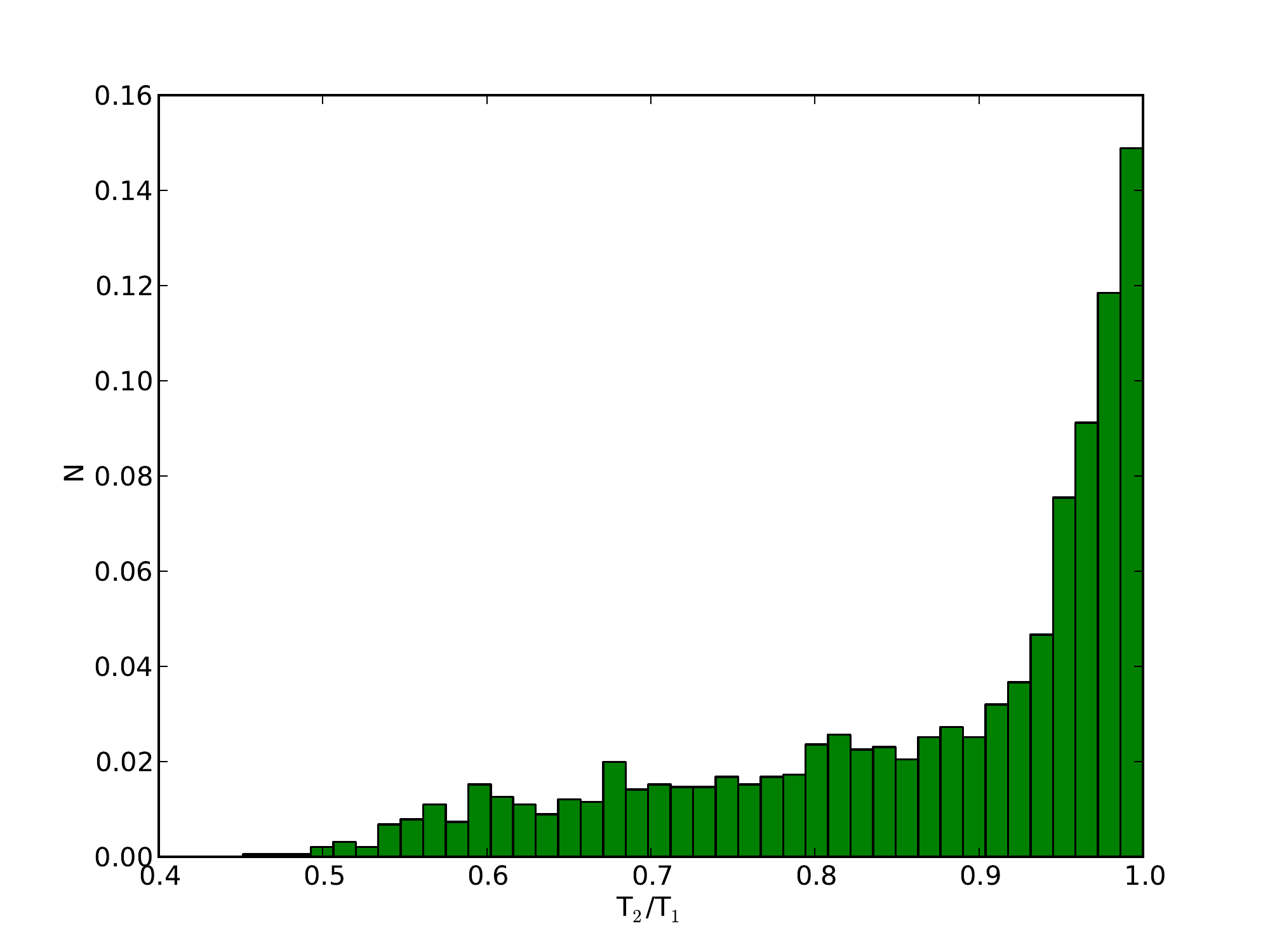}}
\caption{The normalised distribution of $T_2/T_1$ for solar type stars, total number 1908.}
\label{gdist}
\end{figure}

\begin{figure}
\resizebox{\hsize}{!}{\includegraphics{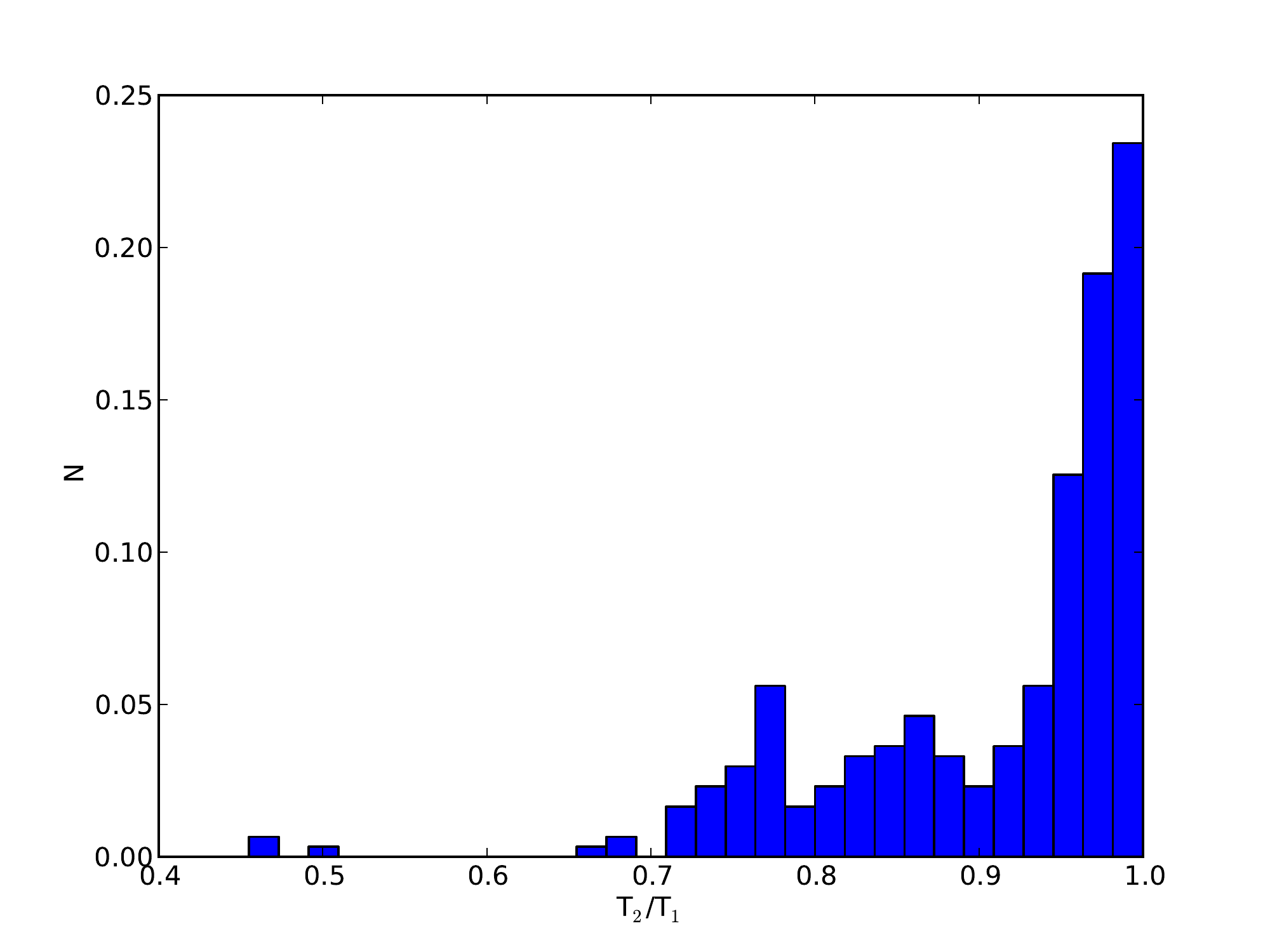}}
\caption{The normalised distribution of $T_2/T_1$ for stars cooler than 5200K, total number 303.}
\label{mdist}
\end{figure}

\begin{figure}
\resizebox{\hsize}{!}{\includegraphics{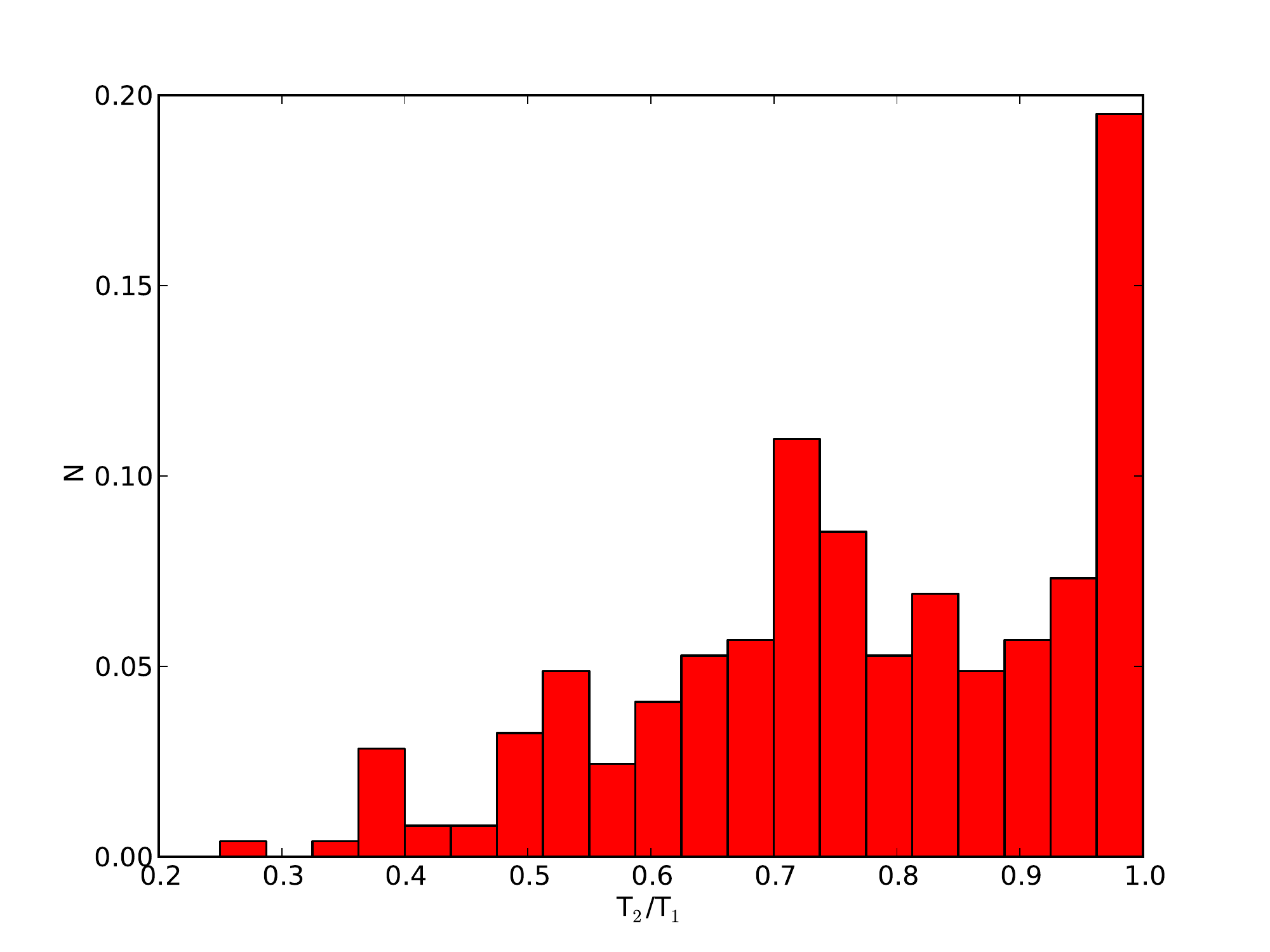}}
\caption{The normalised distribution of $T_2/T_1$ for stars hotter than 7500K, total number 246.}
\label{adist}
\end{figure}

We form distributions of our output parameters using the `final' values as detailed in the results catalogue. Systems for which no good final value could be formed were discarded (this left 2457 of the original 2610 systems, in at least one dataset), While the $R_1/D$ distribution is generally uninformative due to the unknown distance, the $T_1$ and $T_2$ distributions, and in combination the $T_2/T_1$ distribution, is worth noting. The $R_2/R_1$ distribution is in general poorly fit so is again uninformative. The results for $T_1$ and $T_2$ are shown in Figures \ref{t1dist} and \ref{t2dist}. In presenting the temperature ratio, we show the total distribution (Figure \ref{t1t2dist}), and also the distribution split by stellar spectral type. We show the results for `cool' ($T_1<5200K$, roughly K and M stars), solar type ($5200 \le T_1 < 7500K$, roughly F and G stars) and `hot' ($T_1 >= 7500K$, roughly A stars). These are presented normalised to their sample sizes in Figures \ref{gdist}, \ref{mdist}, and \ref{adist}, and are discussed in Section \ref{sectdiscuss}.

\begin{figure}
\resizebox{\hsize}{!}{\includegraphics{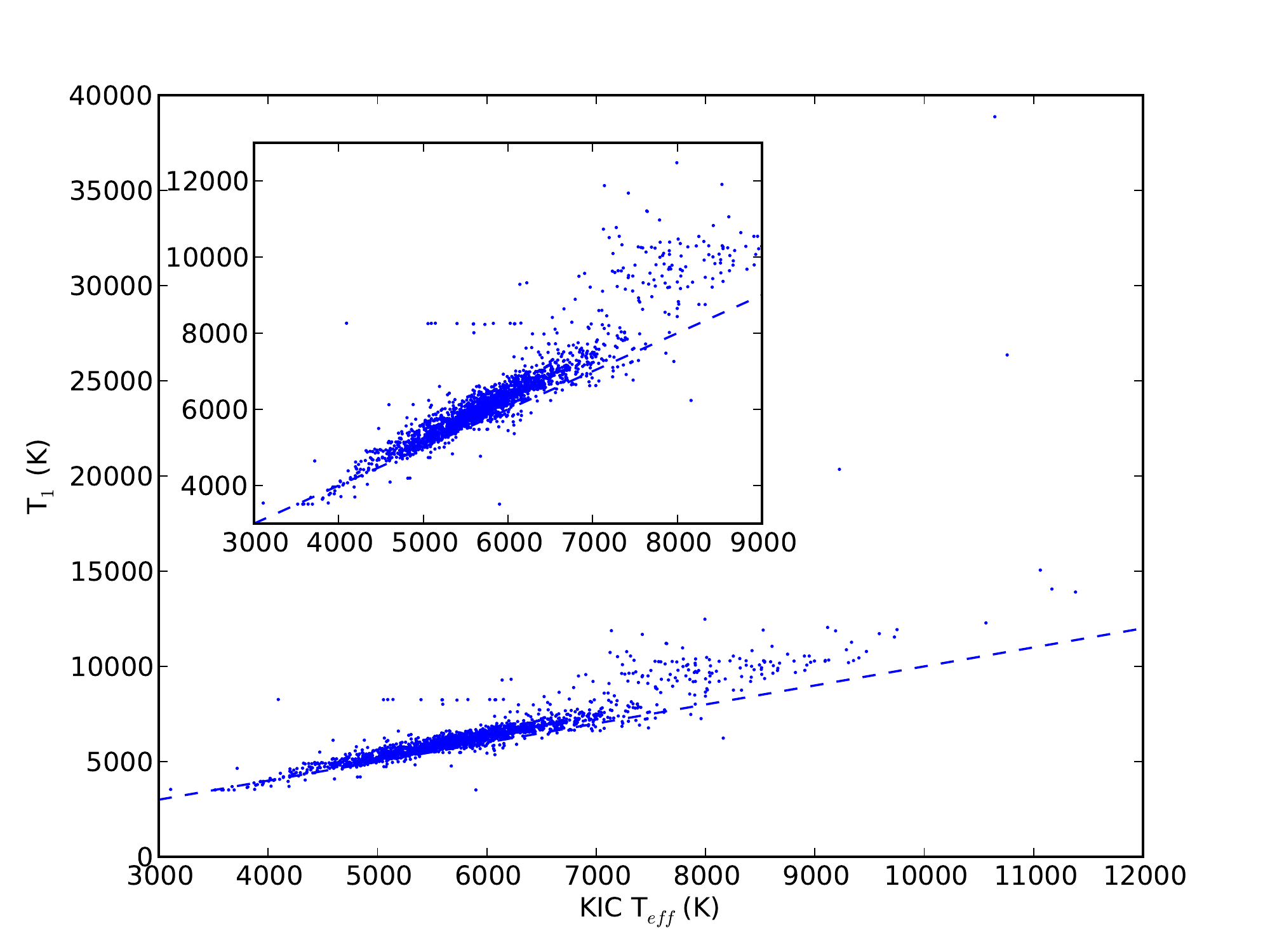}}
\caption{Comparison of KIC $T_{eff}$ and final catalogue $T_1$ values. The dashed line represents a 1:1 match. The difference arises from the inclusion of two stars in our model as compared to the single star of the KIC, and also our higher temperature limit.}
\label{kiccompfig}
\end{figure}

It is also worth comparing our results to the KIC itself. Assuming a single star as the KIC does will tend to focus on the primary star as the dominant source of flux. As such we compare the KIC $T_{eff}$ to our $T_1$ in Figure \ref{kiccompfig}. A general trend of an increase in our temperatures over the KIC's can be seen - this is expected, as in each of these systems an extra contribution from a usually cooler star has been included. The effect of our increased temperature limit over the KIC is also notable.

\section{Discussion}
\label{sectdiscuss}
\subsection{Overview}
We have applied a two stellar component MCMC model to the eclipsing binary stars of the KEBC, using a larger wavelength range than that previously utilised by the KIC. In particular we add the U band, in principle improving our results for hotter stars. We find that the useful results of our model are the effective temperatures of both stars, as well as the parameters $R_1/D$ and $R_2/R_1$. For values of $T_2/T_1$ close to unity, these parameters are not individually fit and the combination $(R_1^2+R_2^2)/D^2$ is instead. The errors on these parameters were found to be on average 370K in $T_1$, 620K in $T_2$, \num{2.5e-12} in $R_1/D$, and 0.23 in $R_2/R_1$. We anticipate that this catalogue will be of use in target selection of Kepler eclipsing binary stars, approximation of parameters and to generally inform anyone wishing to make use of these objects' extremely high-precision light curves. We note that these results use temperature ratios derived from the eclipse depth measurements of the KEBC, and as such they rely on that catalogue to that extent. While the primary star temperatures will only be slightly affected by a largely erroneous temperature ratio, the secondary star temperatures are more sensitive.

\subsection{Shortcomings}
\label{shortcomings}
There are various issues involved in the production of this catalogue, the worst of which we summarise here. Some of these issues are in common with those of the KIC \citep{Brown:2011dr}.

\subsubsection{High/Low T}
The CK stellar atmospheres which we use have a lower limit of 3500K. As such, temperatures which lie close to this value ($T<3750K$) should not be taken as accurate. In terms of selecting cool stars however they can be used - while the temperature is not accurate, any object with a catalogue temperature around this level is unambiguously cool. At high temperatures ($T>9000K$) a large systematic effect, visible in Fig. \ref{t1test}, means that our temperatures have much higher errors, and are likely underestimated. Again, these are still hot stars, but the exact temperature values should not be trusted.

\subsubsection{Binary Solid Angle vs $R_1/D$ and $R_2/R_1$}
As described in Section \ref{secttesting}, the $R_1/D$ and $R_2/R_1$ values are generally poorly constrained. This is because for most of the systems in the KEBC, the temperatures of the two stars are close enough that little information is contained on the secondary star's relative contribution to the flux. Fig. \ref{rrattrattest} shows that only systems with $T_2/T_1\leq 0.6$ constrain $R_1/D$ and $R_2/R_1$, and in these cases there is still a systematic underestimate of \mytilde0.2 in $R_2/R_1$ which has not been adjusted for. In the other systems, which represent the majority of those here, the binary solid angle, equal to $(R_1^2 + R_2^2)/D^2$ and representing a combined measure of the level of flux incoming at the Earth is constrained instead, and can be recovered from the catalogue values as $(R_1/D)^2(1 + (R_2/R_1)^2)$.

\subsubsection{Contamination}
As has been mentioned previously, our model assumes only two stellar components, and does not take account of additional sources of flux. The proportion of KEBC objects with additional companions is \mytilde20\% \citep{Rappaport:2013de}, meaning they represent a significant part of our catalogue. We have removed data flagged as contaminated in the various photometric surveys used, and marked objects which are confirmed as or possible 3+ body systems \citep{Rappaport:2013de,Carter:2011kx,Derekas:2011jsa,Conroy:2013tw,Gies:2012ks}. These systems can still be of use - for example, KIC5897826 is a triple star and is tested successfully against our two star model in Table \ref{tabperformance}. In cases where more than two stellar components are present in a system, the derived physical properties will be affected to the extent that the extra companions contribute to the flux.

\subsubsection{Extinction and Reddening}
Our use of the KIC $A_V$ values allows us to reduce our already large parameter space while still using reasonable extinction values. Attempts to fit these values ourselves were very poorly constrained. Extinction is not generally well constrained - typical errors on the KIC E(B-V) values are of 0.1 mag, implying a 0.31 mag typical error on $A_V$ \citep{Brown:2011dr}, which is of the order of the $A_V$ values. The KIC paper itself notes the problems involved in its extinction parameters, including no account of small scale structure in the interstellar extinction. These errors are incorporated here, and will affect the systems in our catalogue. We have attempted to include the systematic effects of bad extinction in our errors (see Section \ref{secterrors}), but users should be aware that anomalously high extinctions will produce too high temperatures, and the reverse for low extinctions.

\subsection{Distributions}
While our individual stellar temperatures are not constrained by spectroscopy, and so have a larger potential for biases and non-physical effects to show themselves, they offer a particularly large sample size - 303 cool stars, 1908 solar type, and 246 hot stars. However, as the stars under consideration are drawn from an unconstrained sample of ages (and the expected stellar temperature varies with evolution), the temperature distributions are generally uninformative by themselves. The temperature ratio distributions split by spectral type show slight differences, however due to the lower number of samples in each bin of the cool and hot distributions there is not a strong case for any statistically significant difference. This implies that systematic effects in the fitting of different temperature stars were not strong, supporting the robustness of our results across the different temperature regimes. All of the presented distributions have potential biases present, in the effect of the particular combination of colour bands which we utilise, as a residual effect of the KEBC recovery of eclipse depths, or as a sampling effect in the eclipsing binaries which Kepler selects.
 
 \begin{table*}
\caption{Performance on known systems}
\label{tabperformance}
\begin{tabular}{@{}llllllr@{}}
\hline
KIC ID & Description &T1 (Actual, K) & T1 (This Work, K) & T2 (Actual, K) & T2 (This Work, K) & Reference \\
\hline
12644769 & Kepler-16 & $4450\pm150$ & $4013\pm350$ & N/A & $3984\pm550$ & \citet{Doyle:2011ev}\\
8572936 & Kepler-34 & $5913\pm130$ & $5937 \pm 380$ & $5867\pm130$ & $5943 \pm 610$ & \citet{Welsh:2012kl}\\
9837578 & Kepler-35 & $5606\pm150$ & $5902 \pm 370$ & $5202\pm100$ & $4913 \pm 620$ & \citet{Welsh:2012kl}\\
6762829 & Kepler-38 & $5623\pm50$ &$ 5834\pm 350$ & N/A & $ 3539\pm 540$ & \citet{Orosz:2012ip}\\
10020423 & Kepler-47 & $5636\pm100$ & $ 5881\pm 350$ & $3357\pm100$ & $ 3985\pm 680$ & \citet{Orosz:2012ku}\\
4862625 & PH1 & $6407\pm150$ & $ 6705\pm 390$ & $3561\pm150$ & $ 4626\pm 860$ & \citet{Schwamb:2012ts}\\
10661783 & $\delta$-Scuti & $8000\pm160$  & $ 9197\pm 480$ & N/A & $ 7714\pm 810$ & \citet{Southworth:2011ks}\\
5897826 & Triple System & $5875\pm100$  & $ 5798\pm 380$ & N/A & $ 5951\pm 750$ & \citet{Carter:2011kx}\\
6889235 & A+WD binary & $14500\pm500$  & $ 10874\pm 470$& $9500\pm250$ & $ 9519\pm 1000$ & \citet{Bloemen:2012je}\\
7975824 & sdB+WD binary & $34730\pm250$  & $ 32604\pm 2190$ & $15900\pm300$ & $ 10753\pm 1460$ & \citet{Bloemen:2010dc}\\
9472174 & sdB+dm binary & $29564\pm106$  & $ 38868\pm 387$ & N/A & $ 16350\pm 2090$ & \citet{ostensen:2010dr}\\
\hline
\end{tabular}
\end{table*}

 \subsection{Performance on known objects}
 
 A limited number of these eclipsing binary systems, analysed in depth for other reasons, are available to compare with our results. While this sample is far from ideal (by definition these are systems which were selected as `interesting' for a variety of reasons, and hence unlikely to be typical) we would still hope to predict their temperatures reasonably well. We found 11 systems where detailed analysis had been done. Five of these represent interesting stellar objects (e.g. sdB+white dwarves, triple systems) whereas the other 6 were circumbinary planet hosts. In the stellar cases the primary star often dominates the flux, and in these circumstances we would expect to fit this star reasonably well. The spectroscopically derived individual temperatures are compared to ours in Table \ref{tabperformance}. We chose to contain our comparison to temperatures, as in the majority of cases no distance information is available. Encouragingly both the primary and secondary temperatures fit well, within the errors we derive. For the three hot star entries in the table, the fits are generally worse, in all cases lying outside $3\sigma$ for at least one star. They are however, unambiguously high, representing some of the highest temperatures in our catalogue. This highlights that while we can select `hot' stars reasonably well, precise temperatures at values greater than \mytilde9000K can have larger errors. This does not affect the selection of `hot' stars used in the above distributions.

\section*{Acknowledgements}
The authors would like to thank E. Stanway for use of computing resources to complete this work, and D. Steeghs for helpful conversations regarding the Kepler INT Survey. We also thank Andrej Pr\v{s}a for comments which improved the paper.

\bibliography{papers281013}
\bibliographystyle{mn2e_fix}

\appendix
\section{Generation of $T_2/T_1$}
\label{appt2t1gen}
\subsection{Known Parameters}
The KEBC provides periods for each of the binary systems, as well as eclipse depths and measurements in phase of eclipse durations and separations. We produced eccentricity and argument of periapse ($\omega$) measurements using these durations and separations, through the formulae of \citet{kalmilbook}. Initial values of $T_2/T_1$ were produced directly from the ratio of eclipse depths. In some cases no eclipse depth ratio or phase measurements were available, usually due to non-detection of the secondary eclipse.

\subsection{Measurement Scatter}
\label{appmeasscat}
This was taken to be a gaussian error of 0.025 and 0.05 for over contact and non-overcontact binaries respectively, from the test Figures 8 and 10 of \citet{Prsa:2011dx}. Overcontact systems were defined as having morphology parameter greater than 0.7 \citep[see][]{Matijevic:2012di}. While these numbers come from older versions of the KEBC, they represent a reasonable estimate of the uncertainty in recovering $T_2/T_1$ from eclipsing binary light curves.

\subsection{Eccentricity Correction}
This arises from the possibility of different surface areas being occulted in primary and secondary eclipse. The ratio of areas acts as a correction to the temperature ratio, in the form

\begin{equation}
\label{eqnsurfbri}
\frac{T_2}{T_1} = \frac{A_1}{A_2}\left(\frac{D_2}{D_1}\right)^{\frac{1}{4}} 
\end{equation}

\noindent where D represents an eclipse depth and A the stellar surface area occulted in eclipse. The area eclipsed can be calculated from Equation 1 of \citet{Mandel:2002bb}, and is a function of the projected orbital separation at eclipse, plus the stellar radii. The projected orbital separation at eclipse is then itself a function of period, eccentricity, argument of periapse, system mass and inclination. This leaves the radii, mass and inclination unknown. We estimated the possible effect of this correction by drawing 10000 random sample binaries for each KEBC binary, with primary radii drawn uniformly between 0.2 and 1.56 $R_\odot$. 1.56$R_\odot$ is the \citet{Torres:2010eoa} radius of a T=12000K star with surface gravity 4.5 and 0.02 metallicity, chosen as an upper limit as the majority of our catalogue lie below this temperature. Secondary radii were drawn uniformly between 0.2 and the primary radius. Inclinations were constrained such that both stars eclipse. Given these radii, masses were found from the zero-age main sequence, solar metallicity models of \citet{Girardi:2000hj}. The area correction was calculated for each sample, and applied to Equation \ref{eqnsurfbri}, giving a distribution of $T_2/T_1$. We found this to be of the form

\begin{equation}
\label{eqnareadist}
\begin{split}
P(\frac{T_2}{T_1}) \propto & \mspace{8mu} \gamma\delta(\mu) \\ &+ (1-\gamma)\left(\alpha_1e^{\beta_1\lvert \frac{T_2}{T_1}-\mu\rvert} + \alpha_2e^{\beta_2\lvert \frac{T_2}{T_1}-\mu\rvert}\right)
\end{split}
\end{equation}

\noindent where $\mu$ represents the initial value of $T_2/T_1$, produced from Equation \ref{eqnsurfbri} with $A_1/A_2$ set to unity. This consists of two components, a delta function at $\mu$ (from the samples where both eclipses were total, leading to zero correction), and a continuous distribution which we found to be best fit by a sum of exponentials.  The remaining parameters of Equation \ref{eqnareadist} were found for each KEBC binary. We limit ourselves to this first order treatment of the eccentricity effect, ignoring limb darkening and starspots, which are both poorly constrained here.

\subsection{Combined $T_2/T_1$ Distribution}
The above inputs were combined into a final distribution of $T_2/T_1$, suitable for input as a prior into the MCMC. This distribution is found by convolving the measurement and eccentricity correction terms. For simplicity, we convolved the measurement gaussian with the delta function component of the eccentricity correction only. This is equivalent to approximating the exponential terms to dominate outside of approximately one gaussian standard deviation. This led to a final $T_2/T_1$ distribution of the form

\begin{equation}
\label{eqntratdist}
\begin{split}
P(\frac{T_2}{T_1}) =& \mspace{8mu} \frac{1}{N} \Biggl( \gamma\mathcal{N}(\mu,\sigma) \\ &+ (1-\gamma)\left(\alpha_1e^{\beta_1\lvert \frac{T_2}{T_1}-\mu\rvert} + \alpha_2e^{\beta_2\lvert \frac{T_2}{T_1}-\mu\rvert}\right) \Biggr)
\end{split}
\end{equation}

\noindent where N is a normalisation constant and $\sigma$ is set as described in Sect \ref{appmeasscat}. In cases where no eclipse depth ratio or duration information was available, we set the prior on $T_2/T_1$ to be a gaussian with mean 0.7 and standard deviation 0.5, representing a very weak constraint.

\end{document}